\documentclass[12pt]{article}
\usepackage{amsmath,amssymb,amsthm}

\newcommand{\R}{\mathbb{R}}
\newcommand{\cl}[1]{\mathrm{cl}(#1)}
\newcommand{\inte}[1]{\mathrm{int}(#1)}
\newcommand{\bd}[1]{\mathrm{bd}(#1)}
\newcommand{\vac}{V}
\newcommand{\wac}{W}

\newcommand{\ball}{\blacksquare}
\newcommand{\sphere}{\square}
\newcommand{\spheredot}{\boxdot}

\renewcommand{\geq}{\geqslant}
\renewcommand{\emptyset}{\varnothing}
\newcommand{\Aff}{\mathbb A}

\newtheorem*{blanktheorem}{Theorem}
\newtheorem{theorem}{Theorem}
\newtheorem{lemma}{Lemma}

\newtheorem*{fact}{Fact}
\title{Two- versus three-dimensional connectivity testing 
of first-order queries \\ to semi-algebraic sets}
\author{Floris Geerts\thanks{Contact author. Address:
Helsinki Institute for Information Technology, Basic Research
Unit, PO Box 26 (Teollisuuskatu 23), FIN-00014, Finland. Phone:
+358 9 19144037. Fax: +358 9 191 44441. Email:
floris.geerts@cs.helsinki.fi},
Lieven Smits,
and Jan Van den Bussche\thanks{Limburgs Universitair Centrum,
Diepenbeek, Belgium.}}
\date{}

\begin{document}
\maketitle

\begin{abstract}
This paper addresses the question whether one can
determine the connectivity of a semi-algebraic set in three dimensions by
testing the connectivity of a finite number of two-dimensional ``samples'' of
the set, where these samples are defined by first-order queries.
The question is answered negatively for two classes of first-order queries:
cartesian-product-free, and positive one-pass.
\end{abstract}

\section{Introduction}

Semi-algebraic sets provide a useful model for spatial datasets \cite{cdbbook}.
First-order logic over the reals (FO) then provides
a basic query language for expressing queries about such spatial data.
The power of FO, however, is too limited.  In particular, testing whether a
set in $\R^n$ is topologically connected is not expressible in FO
for $n \geq 2$ (for $n=1$ it is easily expressed).

The obvious reaction to this limitation of FO is to enrich it with an explicit
operator for testing connectivity, as proposed by Giannella and Van Gucht
\cite{giann} and by Benedikt et al.~\cite{bgls}.  This operator can be
applied not just to the dataset itself, but also to any set derived from
the original set by an FO query.

The question now arises whether the connectivity of a set in $\R^n$ can
be tested by testing the connectivity of a finite number of
sets in $\R^{n-1}$, constructed from the original set by FO queries.
For $n=2$, the answer is clearly negative, because connectivity in $\R^1$ is
expressible in FO, and therefore a positive answer would imply that
also connectivity in $\R^2$ would be expressible in FO, which we know is not
true.  It is intuitive to conjecture
that the answer is negative for all $n \geq 2$.

While this conjecture in its generality remains open (and seems very hard to
prove), we have proven it for two fragments of FO\@.  In the first fragment,
cartesian product is disallowed.  In the second fragment, negation is
disallowed, and the query must be ``one pass'' in a sense that can be made
precise.  Our treatment of
the second fragment is for $n=3$ only.

\section{Preliminaries}

\paragraph{Semi-algebraic sets}
A \emph{semi-algebraic
set in $\R^n$} is a finite union of sets definable by conditions of
the form $ f_1(\vec{x})=\cdots=f_k(\vec{x})=0$,
$g_1(\vec{x})>0$, \dots, $g_\ell(\vec{x})>0$, with
$\vec{x}=(x_1,\ldots,x_n)\in\R^n$, and where $f_1(\vec{x})$,
\dots, $f_k(\vec{x})$, $g_1(\vec{x})$, \dots,
$g_\ell(\vec{x})$ are multivariate polynomials in the variables
$x_1,\ldots, x_n$ with real coefficients.

Semi-algebraic sets form a very robust class; for example, any set definable
by a formula with quantifiers
in first-order logic
over the reals is semi-algebraic (i.e., definable also without quantifiers;
this is the Tarski--Seidenberg principle \cite{bcr}).

\paragraph{Relational algebra} To express first-order queries about
a set $S$ in $\R^n$, we use not the formalism of first-order logic,
but the equivalent formalism of relational algebra expressions
(RAEs).  These are inductively defined as follows.  The symbol $S$
is a RAE, of \emph{arity} $n$.  Any constant semi-algebraic set in
$\R^k$, for any $k$, is a RAE of arity $k$.  If $e_1$ and $e_2$
are REAs of arities $k_1$ and $k_2$ respectively, then the cartesian
product $(e_1\times e_2)$ is a RAE of arity $k_1+k_2$, and provided
that $k_1=k_2=k$, the union $(e_1\cup e_2)$, the intersection $(e_1 \cap e_2)$
and the difference
$(e_1-e_2)$ are RAEs of arity $k$.  Finally, if $e$ is a RAE of
arity $k$, and $i_1,\ldots,i_p\in\{1,\ldots,k\}$, then the projection
$\pi_{i_1,\ldots,i_p}(e)$ is a RAE of arity $p$.

When applied to a given set $A$ in $\R^n$, a RAE
$e$ of arity $k$ evaluates in the natural way to a set $e(A)$ in $\R^k$.
When $A$ is semi-algebraic, $e(A)$ is too,
by the Tarski--Seidenberg principle.

\paragraph{Notation}
We will use the following notations.
\begin{itemize}
\item
The topological closure
of a set $A\subseteq\R^n$ is denoted by $\cl{A}$, its interior is
denoted by $\inte{A}$ and its boundary $\cl{A}-\inte{A}$ is
denoted by $\bd{A}$.
\item
The $n$-dimensional closed unit ball
centered around the origin is denoted by $\ball$; 
the $n$-dimensional unit sphere centered around the origin by
$\sphere$; and
the union $\sphere \cup \{\vec{0}\}$ by $\spheredot$.
\item
The set of affine transformations from $\R^n$ to $\R^n$ (compositions
of a scaling and a translation) is denoted by $\Aff$.
\end{itemize}

\section{Cartesian-product-free queries}

A RAE is called \emph{cartesian-product-free} if it does not use cartesian
product.  An example of such a RAE is
$$ \pi_{1,2}((S \cap \Gamma_1) \cup (\Gamma_2-S)) - \pi_{1,3}(S
\cup \Gamma_3)$$ where $\Gamma_1$, $\Gamma_2$ and $\Gamma_3$ can
be arbitrary semi-algebraic sets in $\R^3$ and $S$ is ternary (i.e., stands
for a set in $\R^3$).

In this section, we prove that
the connectivity of a semi-algebraic set in $\R^3$ cannot be determined by
sampling it using a finite number of binary
cartesian-product-free RAEs.
\begin{theorem} \label{cartfree}
Let $S$ range over sets in $\R^3$.
For any finite collection $e_1$, \dots, $e_\ell$ of binary
cartesian-product-free RAEs over $S$, there exist two
semi-algebraic sets $A$ and $B$ in $\R^3$ such that
\begin{enumerate}
\item
$A$ is connected;
\item
$B$ is disconnected;
\item
$e_i(A) = e_i(B)$ for $i=1,\dots,\ell$.
\end{enumerate}
\end{theorem}

Toward the proof, we start with the following observation.
\begin{lemma}\label{lem:findset}
Let $\Lambda_0$, $\Lambda_1$, \dots, $\Lambda_k$ be nonempty
semi-algebraic sets in $\R^3$, where $\Lambda_0$ is open. Then
there exists a partition $\{I,J\}$ of
 $\{1,\ldots,k\}$ and an
open semi-algebraic set $\vac\subseteq\Lambda_0$ such that
\begin{itemize}
\item
$\vac\subseteq \Lambda_i$ for  $i \in I$, and
\item
$\vac\cap\Lambda_j=\emptyset$ for $j \in J$.
\end{itemize}
\end{lemma}
\begin{proof}
By induction on $k$.
If $k=0$, set
$I=\{0\}$, $J=\emptyset$, and $\vac=\Lambda_0$.

If $k>0$,
consider the set
$\{\Lambda_0,\Lambda_1,\ldots,\Lambda_{k-1}\}$. Then by the
induction hypothesis, there is a partition $\{I',J'\}$ of
$\{1,\ldots,k-1\}$ and an open set $\vac'\subseteq\Lambda_0$
satisfying the condition as stated in the lemma for $k-1$.  
Since
$\vac'=(\vac'\setminus\Lambda_k)\cup(\vac'\cap\Lambda_k)$, since
$\dim{\vac'}=3$, and since $\dim(A\cup B)=\max\{\dim{A},\dim{B}\}$
for semi-algebraic sets $A$ and $B$, at least one of the following two cases
occurs:
\begin{enumerate}
\item $\dim(\vac'\setminus\Lambda_k)=3$, in which case we choose
$\vac$ an open subset of $\vac'\setminus\Lambda_k$, and set
$I=I'$ and $J=J'\cup\{k\}$.
\item
$\dim(\vac'\cap\Lambda_k)=3$, in which case we choose
$\vac$ an open subset of $\vac'\cap\Lambda_k$, and set
$I=I'\cup\{k\}$ and $J=J'$. \qedhere
\end{enumerate}
\end{proof}

The following lemma is the crucial element in our proof of the theorem.
\begin{lemma}\label{lem:projsimple}
For a given open semi-algebraic set $U\subseteq\R^3$, and any
ternary cartesian-product-free RAE $e$, there exists an open set
$V\subseteq U$ such that $e$ is equivalent to an expression of one of the four
possible forms $$ \Gamma, \quad S, \quad S \cup \Gamma, \quad \Gamma - S $$
on all sets $S \subseteq V$, where $\Gamma$ denotes a constant set in $\R^3$.
Moreover, in the last form, $V$ is included in the interior of\/ $\Gamma$.
\end{lemma}
\begin{proof}
Since both the input $S$ to $e$ and the output of $e$ are ternary, and $e$ is
cartesian-product-free, $e$ must be projection-free as well.
By rewriting $(e_1\cap e_2)$ as $(e_1-(e_2-e_1))$ we can ignore the
intersection operator.
We now proceed by induction on the structure of $e$.
The base cases where $e$ is $S$ or $e$ is constant are already in the right
form.

For the cases $e = (e_1\cup e_2)$ and $e=(e_1
- e_2)$, by induction we can find an open set $V_1 \subseteq U$
such that $e_1$ has one of the four possible forms within
$V_1$, and we can further find an open set $V_2 \subseteq V_1$ such that
$e_2$ has one of the four possible forms within $V_2$.
This means that we have to consider $2\times 4 \times 4$ possibilities
(actually less, as there are symmetries), shown
in Tables \ref{tabelu} and \ref{tabelv}.

\begin{table}
\caption{Proof of Lemma~\protect\ref{lem:projsimple}, possibilities for $e_1
\cup e_2$.} \label{tabelu}
$$ \begin{array}{c|cc}
\cup & S & \Gamma_2 \\
\hline
S & S & S \cup \Gamma_2 \\
\Gamma_1 & S \cup \Gamma_1 & \Gamma_1 \cup \Gamma_2 \\
S \cup \Gamma_1 & S \cup \Gamma_1 & S \cup (\Gamma_1 \cup \Gamma_2) \\
\Gamma_1-S & \Gamma_1 & \{\Gamma_1\cup\Gamma_2, (\Gamma_1\cup\Gamma_2)-S\} \\
\hline
\hline
\cup &  S \cup \Gamma_2 & \Gamma_2 - S \\
\hline
S &  S \cup \Gamma_2 & \Gamma_2 \\
\Gamma_1 &
  S \cup (\Gamma_1 \cup \Gamma_2) &
  \{\Gamma_1\cup\Gamma_2, (\Gamma_1\cup\Gamma_2)-S\} \\
S \cup \Gamma_1 & 
  S \cup (\Gamma_1 \cup \Gamma_2) & S \cup (\Gamma_1 \cup \Gamma_2) \\
\Gamma_1-S &
  S \cup (\Gamma_1 \cup \Gamma_2) & (\Gamma_1\cup\Gamma_2)-S
\end{array} $$
\end{table}

\begin{table}
\caption{Proof of Lemma~\protect\ref{lem:projsimple}, possibilities for $e_1
- e_2$.} \label{tabelv}
$$ \begin{array}{c|cc}
- & S & \Gamma_2 \\
\hline
S & \emptyset & \{\emptyset,S\} \\
\Gamma_1 & \Gamma_1-S & \Gamma_1-\Gamma_2 \\
S \cup \Gamma_1 & \{\Gamma_1,\Gamma_1-S\} &
  \{\Gamma_1-\Gamma_2,(\Gamma_1-\Gamma_2)\cup S\} \\
\Gamma_1 - S & \Gamma_1-S &
  \{\Gamma_1-\Gamma_2, (\Gamma_1-\Gamma_2) - S\} \\
\hline
\hline
- & S \cup \Gamma_2 & \Gamma_2 - S \\
\hline
S & \emptyset & S \\
\Gamma_1 &
  \{\Gamma_1-\Gamma_2, (\Gamma_1-\Gamma_2) - S\} &
  \{\Gamma_1-\Gamma_2, (\Gamma_1-\Gamma_2) \cup S\} \\
S \cup \Gamma_1 &
  \{\Gamma_1-\Gamma_2,(\Gamma_1-\Gamma_2)- S\} &
  S \cup (\Gamma_1-\Gamma_2) \\
\Gamma_1 - S &
  \{\Gamma_1-\Gamma_2, (\Gamma_1-\Gamma_2) - S\} &
  \{\Gamma_1-\Gamma_2, (\Gamma_1-\Gamma_2) - S\}
\end{array} $$
\end{table}

Take, for example,
$e=(\Gamma_1 - S) \cup \Gamma_2$.
By applying
Lemma~\ref{lem:findset} to $\Lambda_0 = V_2$ and $\Lambda_1 = \Gamma_1$,
we get a $V \subseteq V_2$ such that either $V \subseteq \Gamma_1$
or $V \cap \Gamma_1 = \emptyset$.  In the latter case, $e$ is equivalent to
$\Gamma_1\cup\Gamma_2$ within $V$. In the former case, $e$ is equivalent to
$(\Gamma_1\cup\Gamma_2)-S$ within $V$,
and we can always shrink $V$ a bit so that it is
included in the interior
of $\Gamma_1\cup\Gamma_2$, in accordance with the statement of the lemma.
In both cases $e$ is in a desired form.  We summarize this in the
corresponding entry in Table~\ref{tabelu}.  All other entries in the tables
are proven similarly, or are trivial.
\end{proof}

We are now ready for the

\begin{proof}[Proof of Theorem~\ref{cartfree}]
A binary cartesian-product-free RAE $e$ over ternary $S$ can be viewed as an
expression built up, using the operators $\cup$ and $-$,
from binary constant sets and binary projections of
ternary cartesian-product-free RAEs.  If $\pi_{i,j}(c)$ is such a projection
occurring in $e$, we call $c$ a \emph{component} of $e$.

By a series of applications of Lemma~\ref{lem:projsimple}, we can get all
components of all the given binary expressions $e_1$, \dots, $e_\ell$ in one
of the four normal forms mentioned in the lemma.  The first application starts
with $U=\R^3$, and every next application takes as $U$ the $V$ produced by
the previous application.  Within the $V$ produced by the final application,
all components are in normal form.

Choose $\tau\in \Aff$ such that $\tau(\ball)\subset V$, and consider the
sets $A = \tau(\sphere)$ (which is connected) and $B = \tau(\spheredot)$ (which is
disconnected).  Now any binary projection $\pi_{i,j}$ of a component $c$ in
normal form yields the same result whether applied to $A$ or to $B$.  Indeed,
if $c$ is of the form $\Gamma$, $S$, or $S \cup \Gamma$ this is clear; if $c$
is of the form $\Gamma - S$ then we recall that Lemma~\ref{lem:projsimple}
guarantees that $V$ is fully included in the interior of $\Gamma$, so 
$\pi_{i,j}(\Gamma-S) = \pi_{i,j}(\Gamma)$.

We can thus conclude that $e_i(A)=e_i(B)$ for $i=1,\dots,\ell$ as desired.
\end{proof}

For simplicity of exposition, in this section, we have stated and proved
Theorem~\ref{cartfree} in three dimensions only.
However, the argument readily generalizes to prove for any $n>2$ that
the connectivity of a semi-algebraic set in $\R^n$ cannot be determined by
sampling it using a finite number of $n-1$-ary
cartesian-product-free RAEs.

\begin{blanktheorem}
Let $n>2$, and let $S$ range over sets in $\R^n$.
For any finite collection $e_1$, \dots, $e_\ell$ of $n-1$-ary
cartesian-product-free RAEs over $S$, there exist two
semi-algebraic sets $A$ and $B$ in $\R^n$ such that
\begin{enumerate}
\item
$A$ is connected;
\item
$B$ is disconnected;
\item
$e_i(A) = e_i(B)$ for $i=1,\dots,\ell$.
\end{enumerate}
\end{blanktheorem}

\section{Positive one-pass queries}

A RAE is called \emph{positive one-pass} if it does not use the difference
operator, and mentions $S$ only once.  An example is
$$ \pi_{3,5} \big ( \Lambda_1\cup (\Lambda_2 \cap (S \times \R^2)) \big) $$
where $S$ is ternary, and $\Lambda_1$ and $\Lambda_2$
are arbitrary semi-algebraic sets in $\R^5$.
As a matter of fact, this example is very representative, in view of the
following:
\begin{lemma} \label{normalform}
Every binary positive one-pass RAE can be written in the
form $$
\pi_{i_1,i_2} \big ( \Lambda_1\cup (\Lambda_2 \cap (S \times \R^k)) \big). $$
\end{lemma}
More generally, it can be verified by induction that every $p$-ary positive
one-pass RAE can be written in the form of the above lemma, with
$\pi_{i_1,i_2}$ replaced by $\pi_{i_1,\dots,i_p}$.

In this section, we prove that
the connectivity of a semi-algebraic set in $\R^3$ cannot be determined by
sampling it using a finite number of binary
positive one-pass RAEs.
\begin{theorem} \label{posone}
Let $S$ range over sets in $\R^3$.
For any finite collection $e_1$, \dots, $e_\ell$ of binary
positive one-pass RAEs over $S$, there exist two
semi-algebraic sets $A$ and $B$ in $\R^3$ such that
\begin{enumerate}
\item
$A$ is connected;
\item
$B$ is disconnected;
\item
$e_i(A) = e_i(B)$ for $i=1,\dots,\ell$.
\end{enumerate}
\end{theorem}

The following lemma essentially proves the theorem.

\begin{lemma}\label{lem:projexpr}
For a given open semi-algebraic set $U\subseteq\R^3$, any
semi-algebraic sets $\Lambda_1$ and $\Lambda_2$ in $\R^{3+k}$, and
any $i_1,i_2\in\{1,2,3,\allowbreak\ldots,\allowbreak k+3\}$, we can
always find an open set $V\subseteq U$ such that for any
$\tau\in \Aff$ with $\tau(\ball)\subset V$,
$$
\pi_{i_1,i_2}\big(\Lambda_1\cup(\Lambda_2\cap
(\tau(\sphere)\times\R^k))\big)
=\pi_{i_1,i_2}\big(\Lambda_1\cup(\Lambda_2\cap
(\tau(\ball)\times\R^k))\big).
$$
\end{lemma}

Assuming this lemma, we can give the

\begin{proof}[Proof of Theorem~\ref{posone}]
By a series of applications of Lemma~\ref{lem:projexpr}, we obtain a $V$ such
that for any $\tau\in \Aff$ for which $\tau(\ball)\subset V$, we have
$e_i(\tau(\sphere))=e_i(\tau(\ball))$ for $i=1,\dots,\ell$.
Since every $e_i$ is positive
(does not use the difference operator), every $e_i$ is monotone with respect
to the subset order. Hence, $e_i(\tau(\sphere)) \subseteq
e_i(\tau(\spheredot)) \subseteq
e_i(\tau(\ball))$ and thus $e_i(\tau(\sphere)) = e_i(\tau(\spheredot))$.
Taking $A = \tau(\sphere)$ and
$B = \tau(\spheredot)$ thus proves the theorem.
\end{proof}

To prove Lemma~\ref{lem:projexpr} we will use the
regular cell decomposition of semi-algebraic sets, whose definition we recall
next.  A function $f:C\rightarrow\R$, where $C\subseteq\R^n$, is called
\emph{regular} if it is continuous and for each
$i\in\{1,\ldots,n\}$ either strictly increasing, strictly
decreasing, or constant in the $i^{\mathrm{th}}$ coordinate.
(Which of these three cases holds may depend on $i$.)
Also, we call $f$ semi-algebraic if its graph is semi-algebraic.

We define a \emph{regular cell} by induction on the number of dimensions.
Regular cells in $\R$ are singletons $\{a\}$, or open intervals
$(a,b)$, $(-\infty,a)$, or $(a,+\infty)$.
Now assume that $C\subseteq\R^n$ is a regular cell, and
$f,g:C\rightarrow\R$ are regular semi-algebraic functions on $C$,
with $f(\vec{x})<g(\vec{x})$ for all $\vec{x}\in C$. Then the sets
$\{(\vec{x},f(\vec{x}))\mid \vec{x}\in C\}$ and $\{(\vec{x},r)\mid
\vec{x}\in C, f(\vec{x})< r < g(\vec{x})\}$ are regular cells in
$\R^{n+1}$. In the latter case, $f$ can be $-\infty$, and
$g$ can be $\infty$.

A \emph{regular cell decomposition of $\R^n$} is a special kind of
partition of
$\R^n$ into a finite number of regular cells. This is also defined
by induction on $n$. A regular decomposition of $\R$ is just any finite
partition of $\R$ in regular cells.  For $n>1$, a regular cell
decomposition of $\R^n$ is a finite partition
$\{\mathcal{S}_1,\allowbreak\dots,\allowbreak\mathcal{S}_k\}$ of
$\R^n$ in regular cells such that
$\{\pi(\mathcal{S}_1),\ldots,\pi(\mathcal{S}_k)\}$ is a regular
cell decomposition of $\R^{n-1}$. Here, $\pi : (x_1,\dots,x_n)
\mapsto (x_1,\dots,x_{n-1})$ is the natural projection of $\R^n$
onto $\R^{n-1}$.

Let $A$ be a semi-algebraic set in $\R^n$. A regular cell
decomposition of $\R^n$ is said to be \emph{compatible} with $A$
if $A$ is a union of regular cells from this decomposition.

\begin{fact}[\cite{dries}]\label{regdecomp}
For every semi-algebraic set $A$ in $\R^n$ there exists a regular
cell decomposition of $\R^n$ compatible with $A$.
\end{fact}
 
Toward the proof of Lemma~\ref{lem:projexpr},
we start with the following observation.

\begin{lemma}\label{lem:extreme}
Let $A\subseteq\R^3$ be a compact semi-algebraic set and let
$f:A\rightarrow\R$ be a regular function. Then
$\min_{A} f=\min_{\bd{A}} f$ and
$\max_A f=\max_{\bd{A}} f$.
If moreover $\bd{A}$ is
connected, then
$f(\bd A)$ equals the interval $[\min_A f, \allowbreak \max_A f]$.
\end{lemma}

\begin{proof}
Since $A$ is closed, $\bd A \subseteq A$ and thus
$\min_A f \geq \min_{\bd A} f$. To show the reverse
inequality, we need to find for any point in $A
- \bd A$ another point in $\bd A$ with the same or lower $f$-value.
Take a point $(x,y,z) \in A - \bd A$, and shoot a straight ray out of that
point in any direction.  Since $A$ is bounded, the ray will intersect $\bd
A$. Let us focus on the two rays orthogonal to the $xy$ plane.
If $f$ is strictly decreasing in $z$, shoot the ray in
increasing $z$ direction to obtain an intersection point with $\bd A$ with
lower $f$-value as desired.
If $f$ is strictly increasing, follow the converse
direction, and if $f$ is constant, any direction will do to find a point in
$\bd A$ with the same $f$-value.
The equality $\max_A f = \max_{\bd A} f$ is proven in the same way.

Now assume $\bd{A}$ is connected. Choose $\vec{x}_{\max}\in\bd{A}$
with maximal $f$-value, and choose
$\vec{x}_{\min}\in\bd{A}$ with minimal $f$-value.
Since
for semi-algebraic sets, connectivity coincides with path
connectivity~\cite{bcr}, there is a continuous path
$\gamma:[0,1]\rightarrow\bd{A}$ such that
$\gamma(0)=\vec{x}_{\min}$ and $\gamma(1)=\vec{x}_{\max}$. Since
$f$ is continuous, so is the composition
$f\circ\gamma$. Since $[0,1]$ is closed and connected,
$f\circ\gamma([0,1])$ must be a closed and connected set in $\R$, and must
therefore equal the interval
$[\min_A f,\max_A f]$.
\end{proof}

We are now ready to embark on the

\begin{proof}[Proof of Lemma~\ref{lem:projexpr}]
First note that
$\pi_{i_1,i_2}\big(\Lambda_1\cup
(\Lambda_2\cap(S\times\R^k))\big)$
is equivalent to $\pi_{i_1,i_2}(\Lambda_1) \cup
\pi_{i_1,i_2}(\Lambda_2\cap (S\times\R^k))$. So we may focus on
expressions of the form
\begin{equation}
\pi_{i_1,i_2}(\Lambda\cap(S\times\R^k)).\label{expr}
\end{equation}
We only need to prove the inclusion
\begin{equation}\label{eq:toprove}
\pi_{i_1,i_2}(\Lambda\cap(\tau(\ball)\times\R^k))\subseteq
\pi_{i_1,i_2}(\Lambda\cap(\tau(\sphere)\times\R^k)),
\end{equation} the
other direction being trivial.
The proof consists of several cases depending on the indices $i_1,i_2$.

\subsection*{Case 1: $i_1,i_2\in\{1,2,3\}$}
Expression~(\ref{expr}) is equivalent to
$\pi_{i_1,i_2} (E)$, where $E$ is
$\pi_{1,2,3}(\Lambda\cap(S\times\R^k))$.
Applying Lemma~\ref{lem:findset} to
$\Lambda_0=\R^3$ and
$\Lambda_1=\pi_{1,2,3}(\Lambda)$,
we get
an open set $\vac$ such that we are in one of the following two
cases.
\begin{enumerate}
\item $\vac\cap\Lambda_1=\emptyset$.

Within $\vac$, expression $E$,
and hence also~(\ref{expr}),
reduces to the empty set, so the
inclusion~(\ref{eq:toprove}) to be proven trivially holds within $\vac$.
\item $\vac\subset\Lambda_1$.

Within $\vac$, expression $E$ now reduces to $S$, so
expression~(\ref{expr}) reduces to $\pi_{i_1,i_2}(S)$.  In
particular this holds for both $S=\tau(\ball)$ and $S=\tau(\sphere)$,
where $\tau\in \Aff$ such that $\tau(\ball)\subset\vac$.  Since
$\pi_{i_1,i_2}\tau(\ball)=\pi_{i_1,i_2}\tau(\sphere)$, the
inclusion~(\ref{eq:toprove}) holds within $V$.
\end{enumerate}

\subsection*{Case 2: $i_1\in\{1,2,3\}$, $i_2\notin\{1,2,3\}$}
Expression~(\ref{expr}) is now equivalent to
$\pi_{i_1,4} (E)$, where $E$ now is
$\pi_{1,2,3,i_2}(\Lambda\cap(S\times\R^k))$.
Applying Lemma~\ref{lem:findset} to
$\Lambda_0=\R^3$ and $\Lambda_1=\pi_{1,2,3}(\Lambda)$, we get
an open set $\vac^{(0)}$ such that we are in one of the
following cases.
\begin{enumerate}
\item $\vac^{(0)}\cap\Lambda_1=\emptyset$.

Within $\vac^{(0)}$, expression $E$,
and hence also~(\ref{expr}), reduces to the empty set,
so the inclusion~(\ref{eq:toprove}) holds within
$\vac^{(0)}$.
\item $\vac^{(0)}\subset\Lambda_1$.

Within $\vac^{(0)}$, expression $E$ now reduces to
$A\cap(S\times\R)$, with $A=\pi_{1,2,3,i_2}(\Lambda)$.
Consider a regular cell
decomposition of $\R^4$ compatible with $A$, and write the projection
of this decomposition onto $\R^3$ as $\{C_1,\dots,C_\ell\}$.
Applying Lemma~\ref{lem:findset} to 
$\Lambda_0^{(1)}=\vac^{(0)}$, and $\Lambda_i^{(1)}=C_i\cap \vac^{(0)}$
for $i=1,\dots,\ell$, we 
get an open set $\vac^{(1)}\subset\vac^{(0)}$ contained in a unique cell
$C_j$.
Due to our regular cell decomposition, in particular the parts based on $C_j$, 
within $\vac^{(1)}$ the
expression $E = A\cap(S\times\R)$
can now be written as a union of sets of the
form
\begin{align*}
& E_1 = \{(x,y,z,v)\mid (x,y,z)\in S \land v=f(x,y,z)\} \\
\intertext{or}
& E_2 = \{(x,y,z,v)\mid (x,y,z)\in S \land f(x,y,z)<v<g(x,y,z)\},
\end{align*}
where $f$ and $g$ are regular functions.

Assume that $i_1=3$, so that the inclusion~(\ref{eq:toprove}) to be proven 
becomes $\pi_{3,4}(E(\tau(\ball))) \subseteq \pi_{3,4}(E(\tau(\sphere)))$.
The cases $i_1=1$ and $i_1=2$ are analogous.
Since the projection of a union is the union of the projections,
we can restrict attention to the cases $E=E_1$ and $E=E_2$.
\begin{enumerate}
\item $E = E_1$. \label{22a}

Let $\tau\in \Aff$ such that $\tau(\ball)\subset\vac^{(1)}$.
Take an arbitrary element $(z_0,f(x_0,y_0,z_0)) \in \pi_{3,4}(E(\tau(\ball)))$.
Since $\{(x,y,z)\in \tau(\ball) \mid z=z_0\}$ is compact with connected
boundary, we can apply Lemma~\ref{lem:extreme} to obtain $(x_1,y_1,z_0) \in
\tau(\sphere)$ with $f(x_1,y_1,z_0)=f(x_0,y_0,z_0)$. Hence,
$(z_0,f(x_0,y_0,z_0)) \in \pi_{3,4}(E(\tau(\sphere)))$ as desired.
\item $E = E_2$. \label{22b}

By continuity of $f$ and $g$, and because $f<g$,
there exists an open set $\vac^{(2)}\subset\vac^{(1)}$ within which
$f < C < g$ for some constant $C$.
Within $\vac^{(2)}$, we can then break up $E_2$
into three sets
\begin{align*}
B_1&=\{(x,y,z,v)\mid (x,y,z)\in S\land f(x,y,z)<v<C\} \\
B_2&=\{(x,y,z,v)\mid (x,y,z)\in S\land C<v<g(x,y,z)\} \\
B_3&=\{(x,y,z,v)\mid (x,y,z)\in S\land v=C\}
\end{align*}

The set $B_3$ is an instance of case~(\ref{22a}).
We now show that
the set $B_1$ (and, analogously, $B_2$)
can be reduced to that case as well.
Indeed, within $\vac^{(2)}$,
$$
B_1 = \bigcup_{t\in(0,1)} \{(x,y,z,v)\mid (x,y,z)\in S\land
v=tf(x,y,z)+(1-t)C\}.
$$
We now observe that for any $t\in(0,1)$, the function
$tf + (1-t)C$ is regular, so case~(\ref{22a}) applies to
each $t$ individually.
\end{enumerate}
\end{enumerate}

\subsection*{Case 3: $i_1,i_2\notin\{1,2,3\}$}

Expression~(\ref{expr}) is now equivalent to
$\pi_{4,5} (E)$, where $E$ now is
$\pi_{1,2,3,i_1,i_2}(\Lambda\cap(S\times\R^k))$.
Applying, as always, Lemma~\ref{lem:findset} to
$\Lambda_0=\R^3$ and $\Lambda_1=\pi_{1,2,3}(\Lambda)$, we get
an open set $\vac^{(0)}$ such that either
$\vac^{(0)}\cap\Lambda_1=\emptyset$
or
$\vac^{(0)}\subset\Lambda_1$.

If $\vac^{(0)}\cap\Lambda_1=\emptyset$, 
within $\vac^{(0)}$, expression $E$,
and hence also~(\ref{expr}), reduces to the empty set,
so the inclusion~(\ref{eq:toprove}) holds within
$\vac^{(0)}$.

So we can assume that
$\vac^{(0)}\subset\Lambda_1$.
Within $\vac^{(0)}$, expression $E$ now reduces to
$A\cap(S\times\R^2)$, with $A=\pi_{1,2,3,i_1,i_2}(\Lambda)$.
Consider a regular cell
decomposition of $\R^5$ compatible with $A$, and write the projection
of this decomposition onto $\R^3$ as $\{C_1,\dots,C_\ell\}$.
Applying Lemma~\ref{lem:findset} to 
$\Lambda_0^{(1)}=\vac^{(0)}$, and $\Lambda_i^{(1)}=C_i\cap \vac^{(0)}$
for $i=1,\dots,\ell$, we 
get an open set $\vac^{(1)}\subset\vac^{(0)}$ contained in a unique cell
$C_j$.
Due to our regular cell decomposition, in particular the parts based on $C_j$, 
within $\vac^{(1)}$ the
expression $E = A\cap(S\times\R^2)$
can now be written as a union of sets of the
form
\begin{tabbing}
$E_1 = \{(x,y,z,u,v)\mid (x,y,z)\in S\land
u=f(x,y,z) \land v=g(x,y,z,u)\}$,
\\
$E_2 =
\{(x,y,z,u,v)\mid (x,y,z)\in S$\=${}\land u=f(x,y,z)$ \\
\>${}\land g_1(x,y,z,u)< v <g_2(x,y,z,u)\}$,
\\
$E_3 =
\{(x,y,z,u,v)\mid (x,y,z)\in S$\=${}\land f_1(x,y,z)<u<f_2(x,y,z)$ \\
\>${}\land v=g(x,y,z,u)\}$, or \\
$E_4 =
\{(x,y,z,u,v)\mid (x,y,z)\in S$\=${}\land f_1(x,y,z)<u<f_2(x,y,z)$ \\
\>${}\land g_1(x,y,z,u)<v<g_2(x,y,z,u)\}$,
\end{tabbing}
where $f$, $f_1$, $f_2$, $g$, $g_1$, and $g_2$ are regular functions.

We need to prove $\pi_{4,5}(E(\tau(\ball))) \subseteq 
\pi_{4,5}(E(\tau(\sphere)))$.
Since the projection of an union is the union of the projections,
we can restrict attention to the cases $E=E_1$, $E=E_2$, $E=E_3$, and $E=E_4$.
\begin{enumerate}
\item $E=E_1$. \label{31}
\begin{enumerate}
\item
If $f$ is constant, with value $u_0$, $\pi_{4,5}(E_1)$ reduces to
$$ \{(u_0,g(x,y,z,u_0)) \mid (x,y,z) \in S\} $$ which can be handled by the
same reasoning as in Case~2,~(\ref{22a}).
\item \label{31b}
Now assume $f$ is not constant in $x$; the cases $y$ and $z$ are
analogous.  Look at the projection
$\pi_{2,3,4,5}(E_1)$:
\begin{equation*}
\{(y,z,u,v)\mid \exists x((x,y,z)\in S \land u=f(x,y,z)\land
v=g(x,y,z,u))\}.
\end{equation*}
This set can be written as
\begin{equation}
E'_1 = \{(y,z,u,v)\mid (y,z,u)\in h(S)\land v=k(y,z,u)\},
\end{equation}
where
$h:(x,y,z)\mapsto (y,z,f(x,y,z))$, and
$$k:(y,z,u)\mapsto g(h_x^{-1}(y,z,u),y,z,u), $$ where
$h_x^{-1}$ is the function defined by
$h(h_x^{-1}(y,z,u),y,z)=(y,z,u)$. This
inverse function exists; in fact, because $f$ is regular and non-constant in
$x$, $h$ is a homeomorphism within $\vac^{(1)}$.

Within $\wac^{(0)}=h(\vac^{(1)})$, we
can find an open set $\wac^{(1)}\subset\wac^{(0)}$ such that $k$
is regular. Since $h$ is a homeomorphism, we also can
find an open set $\vac^{(2)}\subset\vac^{(1)}$ such that
$h(\vac^{(2)})\subset\wac^{(1)}$.
Since $\pi_{4,5}(E_1)$ reduces to $\pi_{3,4}(E'_1)$,
we can, within $\vac^{(2)}$,
now again reason analogously as in Case~2,~(\ref{22a}).
\end{enumerate}

\item $E=E_2$. \label{32}

By continuity of $f$, $g_1$ and $g_2$, and because $g_1 < g_2$, 
we can find
an open set $\vac^{(2)}\subseteq \vac^{(1)}$ within which
$$g_1(x,y,z,f(x,y,z))< C <g_2(x,y,z,f(x,y,z))$$ for some constant $C$.
Now reason repeatedly as in case~(\ref{31}) on the following sets:
\begin{tabbing}
$B_1=\{(x,y,z,u,v)\mid (x,y,z)\in S\land u=f(x,y,z) \land v=g_1(x,y,z,u)\}$
\\
$B_2=\{(x,y,z,u,v)\mid (x,y,z)\in S\land u=f(x,y,z) \land v=g_2(x,y,z,u)\}$
\\
$B_3=\{(x,y,z,u,v)\mid (x,y,z)\in S\land u=f(x,y,z) \land v=C)$
\end{tabbing}
We thus obtain $\vac^{(3)} \subseteq \vac^{(2)}$ within which
$\pi_{4,5}(B_i(\tau(\ball))) \subseteq \pi_{4,5}(B_i(\tau(\sphere)))$ for
$i=1,2,3$.
We then break up $E_2$ in $B_3 \cup B_4 \cup B_5$, where
\begin{tabbing}
$B_4=\{(x,y,z,u,v)\mid (x,y,z)\in S\land u=f(x,y,z) \land
g_1(x,y,z,u)< v < C \}$
\\
$B_5=\{(x,y,z,u,v)\mid (x,y,z)\in S\land
u=f(x,y,z) \land C< v < g_2(x,y,z,u) \}$
\end{tabbing}

It remains to treat $B_4$ and $B_5$, and we do this as follows,
similarly to what we did in Case~2,~(\ref{22b}).
Reasoning as in case~(\ref{31}), we have written $\pi_{4,5}(B_1)$ as
$\pi_{3,4}(\{(y,z,u,v)\mid (y,z,u)\in
h(S)\land u=k(y,z,u)\})$. We then can write $\pi_{4,5}(B_4)$ as
$$
\bigcup_{t\in(0,1)}\pi_{3,4}(\{(y,z,u,v)\mid
(y,z,u)\in h(S)\land v=tk(y,z,u)+(1-t)C\}).
$$
We now observe
that for any $t\in(0,1)$, the
function  $t k + (1-t)C$ is regular, so we can reason analogously as in
Case~2,~(\ref{22a})
for each $t\in(0,1)$ individually.  We treat $B_5$ in the same way, now using
the $h$ and $k$ from $B_2$.

\item $E=E_3$. \label{33}

We begin again by determining an open set
$\vac^{(2)}\subset\vac^{(1)}$ within which $f_1 < C < f_2$
for some constant $C$,
and break up $E_3$ in the following sets:
\begin{tabbing}
$B_1=\{(x,y,z,u,v)\mid (x,y,z)\in S\land f_1(x,y,z)<u<C\land v=g(x,y,z,u)\}$
\\
$B_2=\{(x,y,z,u,v)\mid (x,y,z)\in S\land C<u<f_2(x,y,z)\land
v=g(x,y,z,u)\}$
\\
$B_3=\{(x,y,z,u,v)\mid (x,y,z)\in S\land u=C
\land v=g(x,y,z,u)\}$
\end{tabbing}
On $B_3$ we can reason as in case~(\ref{31}) and obtain
an open set
$\vac^{(3)}\subset\vac^{(2)}$ within which $\pi_{4,5}(B_3(\tau(\ball)))
\subseteq \pi_{4,5}(B_3(\tau(\sphere)))$.

We show how to treat $B_1$;  the treatment of $B_2$ is analogous.
Within a certain open set $\vac$ to be determined, we are going to break up
$B_1$ in a special way in two overlapping parts of the following form:
\begin{tabbing}
$\displaystyle
B_{1,1}=\bigcup_{t\in(0,\delta)} \{(x,y,z,u,v)\mid
\begin{aligned}[t] (x,y,z)\in S
& \land u=f(x,y,z)+t \\
& \land v=g(x,y,z,u)\} \end{aligned}$
\\
$\displaystyle
B_{1,2}=\bigcup_{t\in(c_{\vac},C)} \{(x,y,z,u,v)\mid (x,y,z)\in
S\land u=t\land v=g(x,y,z,u)\}$
\end{tabbing}
for certain $\delta$ and $c_{\vac}$, which we are now going to
define.

If $f$ is constant, then $\delta:=0$, and $c_V$ is the
constant value of $f$.

So, suppose that $f$ is not constant in $x$; the cases $y$ and $z$ are
analogous.  Then
$h_t:(x,y,z)\mapsto (y,z,f(x,y,z)+t)$ is a homeomorphism for
every $t$. Let
$$k: (y,z,u,t) \mapsto g((h_t)_x^{-1}(y,z,u),y,z,u),$$
where $(h_t)_x^{-1}$ is the function defined by
$h_t((h_t)_x^{-1}(y,z,u),y,z)=u$.

We now want to find a $\delta$ such that $k_t : (y,z,u) \mapsto k(y,z,u,t)$
is regular for every
$t\in(0,\delta)$. Thereto, consider the (semi-algebraic) set
$$
D=\{(y,z,u,t)\mid (y,z,u)\in h_0(\vac^{(3)}) \land 0<t<1 \land
\frac{\partial k}{\partial y}(y,z,u,t)=0 \}
$$
Using a cell decomposition of $\R^4$ compatible with $D$, we can
find an open set $\wac^{(0)}\subseteq h_0(\vac^{(3)})$ and
a $\delta^{(0)}>0$ such that on $\wac^{(0)}\times(0,\delta^{(0)})$
either $\frac{\partial k}{\partial y}=0$,
i.e., $k$ is constant in $y$,
or $\frac{\partial k}{\partial y}\neq 0$, i.e., $k$ is strictly
monotone in $y$.
Proceeding similarly, we can find $\wac^{(2)} \subseteq \wac^{(1)} \subseteq
\wac^{(0)}$ and $0<\delta^{(2)}<\delta^{(1)}<\delta^{(0)}$ such that 
$k$ is either constant or strictly monotone in $z$
on $\wac^{(1)}\times(0,\delta^{(1)})$, and 
$k$ is either constant or strictly monotone in $u$
on $\wac^{(2)}\times(0,\delta^{(2)})$.  Hence, within $\wac^{(2)}$,
$k_t$ is regular for every $t \in (0,\delta^{(2)})$.

Next, choose an open set $\vac^{(4)}\subset\vac^{(3)}$ and
$0<\delta^{(3)}<\delta^{(2)}$ such that
$h_t(\vac^{(4)})\subset\wac^{(2)}$ for every $t\in(0,\delta^{(3)})$.
We then restrict $\vac^{(4)}$ even further to an open set
$\vac^{(5)}$, and simultaneously choose $\delta^{(4)}$
such that the following conditions are satisfied:
\begin{gather*}
C-\sup_{\vac^{(5)}}f>\delta^{(4)}>0
\\
\sup_{\vac^{(5)}}f - \inf_{\vac^{(5)}}f <
\min\{\delta^{(3)},\delta^{(4)}\}
\end{gather*}
It is now clear that, within $\vac:=\vac^{(5)}$,
we have $B_1 = B_{1,1} \cup B_{1,2}$ where
we put $c_{\vac}:=\sup_{\vac^{(5)}}f$ and
$\delta:=\min\{\delta^{(3)},\delta^{(4)}\}$.

It remains to deal with $B_{1,1}$ and $B_{1,2}$, but this poses no longer any
problems:
\begin{description}
\item[$B_{1,1}$:]
By construction, $k(y,z,u,t)$ is regular for every $t\in(0,\delta)$.
This implies that we can work with the sets
$$     
\{(y,z,u,v)\mid (y,z,u)\in h_t(S) \land v=k(y,z,u,t)\}
$$
as in case~(\ref{31}).
\item[$B_{1,2}$:]
Here, for every $t$, we are back in Case~2,~(\ref{22a}).
\end{description}
\item $E = E_4$.
We begin again by determining an open set
$\vac^{(2)}\subset\vac^{(1)}$ within which $f_1 < C < f_2$
for some constant $C$,
and break up $E_4$ in the following sets:
\begin{tabbing}
$B_1=\{(x,y,z,u,v)\mid (x,y,z)\in S$\=${} \land f_1(x,y,z)<u<C$ \\
\>${} \land g_1(x,y,z,u)<v<g_2(x,y,z,u)\}$
\\
$B_2=\{(x,y,z,u,v)\mid (x,y,z)\in S$\=${} \land C<u<f_2(x,y,z)$ \\
\>${} \land g_1(x,y,z,u)<v<g_2(x,y,z,u)\}$
\\
$B_3=\{(x,y,z,u,v)\mid (x,y,z)\in S$\=${} \land u=C$ \\
\>${} \land g_1(x,y,z,u)<v<g_2(x,y,z,u)\}$
\end{tabbing}
On $B_3$ we can reason as in case~(\ref{32}) and obtain
an open set
$\vac^{(3)}\subset\vac^{(2)}$ within which $\pi_{4,5}(B_3(\tau(\ball)))
\subseteq \pi_{4,5}(B_3(\tau(\sphere)))$.

We show how to treat $B_1$; the treatment of $B_2$ is analogous.
By the same procedure as in case~(\ref{33}), but now working with two
functions $k_1$ and $k_2$ (one for $g_1$ and one for $g_2$), we break up
$B_1$ within a certain open set $\vac$:
\begin{align*}
& B_{1,1}=\bigcup_{t\in(0,\delta)} \{(x,y,z,u,v)\mid
\begin{aligned}[t] & (x,y,z)\in S \land u=f(x,y,z)+t \\
& \quad \land g_1(x,y,z,u)<v<g_2(x,y,z,u)\} \end{aligned}
\\
& B_{1,2}=\bigcup_{t\in(c_{\vac},C)} \{(x,y,z,u,v)\mid
\begin{aligned}[t] & (x,y,z)\in S \land u=t \\
& \quad \land g_1(x,y,z,u)<v<g_2(x,y,z,u)\} \end{aligned}
\end{align*}
We finally deal with $B_{1,1}$ and $B_{1,2}$ as follows:
\begin{description}
\item[$B_{1,1}$:]
By construction, $(k_1)_t$ and $(k_2)_t$ are regular for every
$t\in(0,\delta)$.  Writing $\pi_{4,5}(B_{1,1})$ as
$$
\bigcup_{t \in (0,\delta)} \pi_{3,4}(\{(y,z,u,v) \mid
\begin{aligned}[t] & (y,z,u) \in h_t(S) \\
& \quad \land k_1(y,z,u,t) < v < k_2(y,z,u,t)\}) \end{aligned}
$$
we can therefore reason analogously as in Case~2,~(\ref{22b}) for every $t$
individually.
\item[$B_{1,2}$:]
Here, for every $t$ individually, we are straight back in Case~2,~(\ref{22b}).
\end{description}
\end{enumerate}
The proof of Lemma~\ref{lem:projexpr} is complete.
\end{proof}

\section{Concluding remarks}

We have treated the positive-one pass queries for three-dimensional datasets
only.  Our proof uses only fairly
elementary mathematics.  By using more heavy machinery, one can probably prove
our Theorem~\ref{posone} in general for $n$-dimensional datasets and $n-1$-ary
queries.  Conceivably this generalisation can also be performed starting from
our own proof, but that will be exceedingly laborious.

Extending our proof technique to larger classes of RAEs is not obvious to us.
For instance, when relaxing the one-pass restriction, it is not clear how to
find a good $\tau\in \Aff$ such that $\tau(\ball) \times \tau(\ball)$ is nicely
located. When negation is allowed, the normal form of Lemma~\ref{normalform}
becomes much more complex, with consequences for the case analysis.

Ultimately, one can even go further than the problem posed in the
Introduction, and throw in
connectivity testing of \emph{parameterized} queries, which can then even be
nested \cite{bgls,giann}.

\end{document}